\documentclass[twocolumn,showpacs,superscriptaddress,preprintnumbers,amsmath,amssymb,aps,prl,10pt]{revtex4-1}
\usepackage{graphicx}
\usepackage{bm}
\usepackage{multirow}

\newcommand{\cref}[1]{(\ref{#1})}
\newcommand{\rcite}[1]{[\onlinecite{#1}]}

\begin{document}
\title{Dense Regular Packings of Irregular Non-Convex Particles}

\author{Joost de Graaf}
\affiliation{%
Soft Condensed Matter, Debye Institute for Nanomaterials Science, Utrecht University, Princetonplein 5, 3584 CC Utrecht, The Netherlands
}%
\author{Ren\'e van Roij}%
\affiliation{%
Institute for Theoretical Physics, Utrecht
University, \\ Leuvenlaan 4, 3584 CE Utrecht, The Netherlands
}%
\author{Marjolein Dijkstra}
\affiliation{%
Soft Condensed Matter, Debye Institute for Nanomaterials Science, Utrecht University, Princetonplein 5, 3584 CC Utrecht, The Netherlands
}%

\begin{abstract}
We present a new numerical scheme to study systems of non-convex, irregular, and punctured particles in an efficient manner. We employ this method to analyze regular packings of odd-shaped bodies, not only from a nanoparticle but also both from a computational geometry perspective. Besides determining close-packed structures for many shapes, we also discover a new denser configuration for Truncated Tetrahedra. Moreover, we consider recently synthesized nanoparticles and colloids, where we focus on the excluded volume interactions, to show the applicability of our method in the investigation of their crystal structures and phase behavior. Extensions to the presented scheme include the incorporation of soft particle-particle interactions, the study of quasicrystalline systems, and random packings.
\end{abstract}


\date{\today}

\maketitle

The synthesis of colloids and nanoparticles has advanced enormously over the last decade~\cite{Sun,Manna1,Zhao,Solomon_Glotzer,Quilliet,Manna0}. Currently it is not only possible to synthesize spherical particles, but also a wide variety of convex faceted shapes, such as tetrahedra, cubes and octahedra~\cite{Sun,Solomon_Glotzer}. Perhaps the most remarkable advancement in synthesis techniques is the capability to create with high precision and reproducibility non-convex, irregular, and even punctured particles, e.g., colloidal caps~\cite{Quilliet}, tetrapods~\cite{Manna1}, octapods~\cite{Manna0}, and nanostars~\cite{Zhao}. Along with the increased availability of complex shapes, there is the concurrent increase in the study of their self-assembly into liquid~\cite{Bernal}, amorphous~\cite{Torquato_4,Donev}, and ordered phases~\cite{Chaikin}.  Especially the ordered (quasi)crystalline structures~\cite{Chaikin,Bechinger,Talapin,Akbari_Glotzer} of odd-shaped particles, as well as their material properties, have received a lot of attention in the past years. Interestingly, the research into dense packings of particles is not restricted to materials science, as it is also connected to fields as diverse as discrete geometry and number theory~\cite{Kuperberg,Betke,Hales1}, computer science and biophysics~\cite{Liang}.

 Predictions obtained from computer simulations on the phase behaviour and the self-assembled structures of these particles have been essential in guiding experimental studies and in answering fundamental mathematical questions on the packing of particles. Convex objects such as spheres~\cite{Hales1,Bernal} and ellipsoids~\cite{Donev}, as well as (semi)regular~\cite{Kuperberg,Betke,Schilling,Akbari_Glotzer,Torquato_1,Torquato_2,Chen_Glotzer} and space-filling~\cite{Agarwal} solids have been the subject of intense ongoing investigation. However, ordered structures comprised of {\em  irregular non-convex} particles have hardly been studied by simulation because of the numerical challenges in implementing excluded volume interactions for such systems, due to the complex particle shape and the additional rotational degrees of freedom. Only recently were the first attempts made to study such systems, namely for superdisks and superballs~\cite{Torquato_5}.

In this Letter, we present a novel numerical method to study systems of non-convex irregular polytopes, colloids, and nanoparticles. We employ this method to establish a rigorous lower bound $\phi_{\mathrm{lb}}$ to the volume fraction of their densest regular packing and to predict candidate crystal structures. Our numerical scheme elegantly and systematically reduces simulation studies of these complex systems in a two-fold way. Firstly, the problem of determining overlaps between irregular particles is reduced tremendously, especially when these particles are faceted~\cite{Solomon_Glotzer}, by approximating the particle shape with triangles~\cite{Graaf_Roij1}, since triangle intersections can be straightforwardly determined. Secondly, predicting candidate crystal structures is greatly simplified by using the Floppy Box Monte Carlo method~\cite{Filion} (FBMC), which efficiently analyzes the regular structure on a unit-cell level.  We anticipate our investigation to be the starting point of an exploration of (dis)ordered packings of arbitrarily shaped particles, which rivals the current research effort on more conventional bodies.

To prove the accuracy of our method, we have reproduced the lower bound $\phi_{\mathrm{lb}}$ both for the Platonic and for the Archimedean Solids, see Ref.~\cite{EPAPS}. We find excellent agreement with Refs.~\rcite{Chen_Glotzer,Betke,Torquato_1,Torquato_2}. The deviation between $\phi_{\mathrm{lb}}$ and the literature value $\phi_{\mathrm{lb}}^{*}$ ranges from $2\cdot 10^{-7}$ to $3\cdot 10^{-3}$. For all results obtained by FBMC $\phi_{\mathrm{lb}} < \phi_{\mathrm{lb}}^{*}$, because compression to the mathematically derived packing fraction~\cite{Betke,Chen_Glotzer} is virtually impossible. However, FBMC performs remarkably well for these convex shapes: the simulations we performed typically yielded a very narrow distribution of crystal-structure candidates near the closest-packed configuration, the densest of which only required minimal additional compression to achieve the given value of $\phi_{\mathrm{lb}}$. Moreover, for Truncated Tetrahedra we have discovered a new crystal structure, a dimer lattice, with $\phi_{\mathrm{lb}}=0.988\dots$~\cite{EPAPS}. This is not only mathematically interesting, it is also relevant to the study of nanoparticle systems, since Truncated Tetrahedral particles have recently been synthesized~\cite{Sun}.

\begin{figure}[!htb]
\includegraphics[width=3.375in]{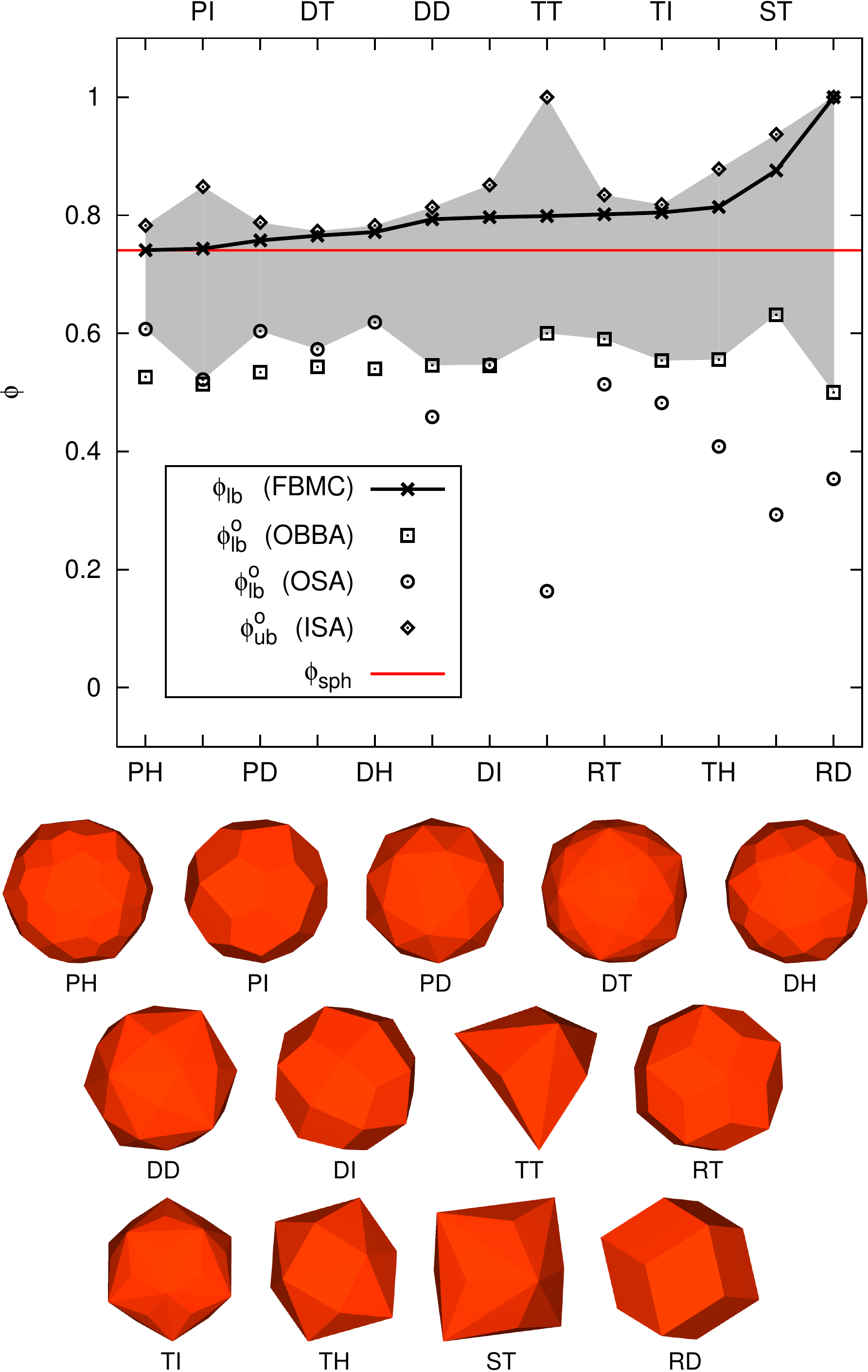}
\caption{\label{fig:cata} \textbf{Upper and lower bounds to the densest packing fraction for the 13 Catalan Solids.} The graph shows the results for the outscribed-sphere approximation (OSA, $\phi_{\mathrm{lb}}^{\circ}$, circles), oriented-bounding-box approximation (OBBA, $\phi_{\mathrm{lb}}^{\circ}$, squares), and FBMC method ($\phi_{\mathrm{lb}}$, connected crosses), which give lower bounds to the maximum packing fraction, and the inscribed-sphere approximation~\cite{Torquato_1} (ISA, $\phi_{\mathrm{ub}}^{\circ}$, diamonds), which gives an upper bound to this packing fraction. The value of the densest packing for spheres $\phi_{\mathrm{sph}}$ is indicated by a red line. A visual representation of the Catalan Solids is given below the graph, the abbreviations are as follows: Pentagonal Hexecontrahedron (PH), Pentagonal Icositetrahedron (PI), Pentakis Dodecahedron (PD), Disdyakis Triacontrahedron (DT), Deltoidal Hexecontrahedron (DH), Disdyakis Dodecahedon (DD), Deltoidal Icositetrahedron (DI), Triakis Tetrahedron (TT) , Rhombic Triacontrahedron (RT), Triakis Icosahedron (TI), Tetrakis Hexahedron (TH), Small Triakis Octahedron (ST), and Rhombic Dodecahedron (RD). Note the improvement of FBMC with respect to the OSA and OBBA lower bound: the maximum packing fraction can henceforth be located in the gray area above the FBMC line.}
\end{figure}

For all prisms and antiprisms with regular $n$-gonal bases ($n > 3$), for all Johnson Solids, and for all Catalan Solids, we have verified Ulam's conjecture~\cite{Gardner}, which states that all convex objects pack denser than spheres. More importantly, for all of these sets we also find that the obtained densest packing for centrally symmetric polyhedra is a Bravais lattice packing, in accordance with the conjecture of Ref.~\rcite{Torquato_1}, and that for the convex, congruent solids without central symmetry it is {\em not} a Bravais lattice, in accordance with the conjecture of Ref.~\rcite{Torquato_2}. Here, we present only the data on the Catalan Solids, which are dual to the Archimedean Solids. In Fig.~\ref{fig:cata} several bounds to the densest packing fraction for the Catalan Solids are shown. The outscribed-sphere approximation (OSA) and oriented-bounding-box approximation (OBBA) yield lower bounds to the maximum packing fraction, the inscribed-sphere approximation~\cite{Torquato_1} (ISA) provides an upper bound. Ref.~\cite{EPAPS} gives a more comprehensive account of the terminology and the employed techniques. It also lists the values given in Fig.~\ref{fig:cata} with 6 digit precision. The ISA in combination with the FBMC result gives a much narrower region in which the densest packing can be found for this group of solids than was previously established. The narrowness of the region can be attributed to the relatively low asphericity of the Catalan Solids.

Other particle shapes for which we have determined the densest-packing crystal structures include the regular Penta- and Heptaprism: $\phi_{\mathrm{lb}} = 0.921\dots$ and $\phi_{\mathrm{lb}} = 0.892\dots$,  respectively~\cite{EPAPS}. The interest in Pentaprisms stems not only from the fact that such particles can nowadays be synthesized~\cite{Geng}, but also from the fact that Pentagons offer the exciting possibility of a structure with quasicrystalline order in two dimensions~\cite{Bauer}. Hence, we consider the prismatic equivalent of Pentagons, Pentaprisms, in our three dimensional (3D) method. Heptaprisms are studied for the purpose of comparison. The structures we obtain by FBMC for these Penta- and Heptaprisms are the 3D columnar continuation of the two dimensional (2D) double-lattices found for Pentagons and Heptagons, respectively. The prisms in different columns, which themselves form a 2D crystal, are out of register. The 2D results, to which we compare our 3D crystals, were established by mathematical techniques~\cite{Kuperberg} and simulations~\cite{Schilling}. Our results indicate that a 3D columnar continuation of a Pentaprism quasicrystalline configuration may also exist, since columnar continuations appear to be common for non-space-filling prismatic systems. For space-filling Tri- and Hexaprisms (and Cubes) we observe both columnar continuations and randomly stacked layers, i.e., the prisms form crystalline planes, but the ordering in different planes is not in register.

\begin{figure*}[!htb]
\includegraphics[scale=0.60]{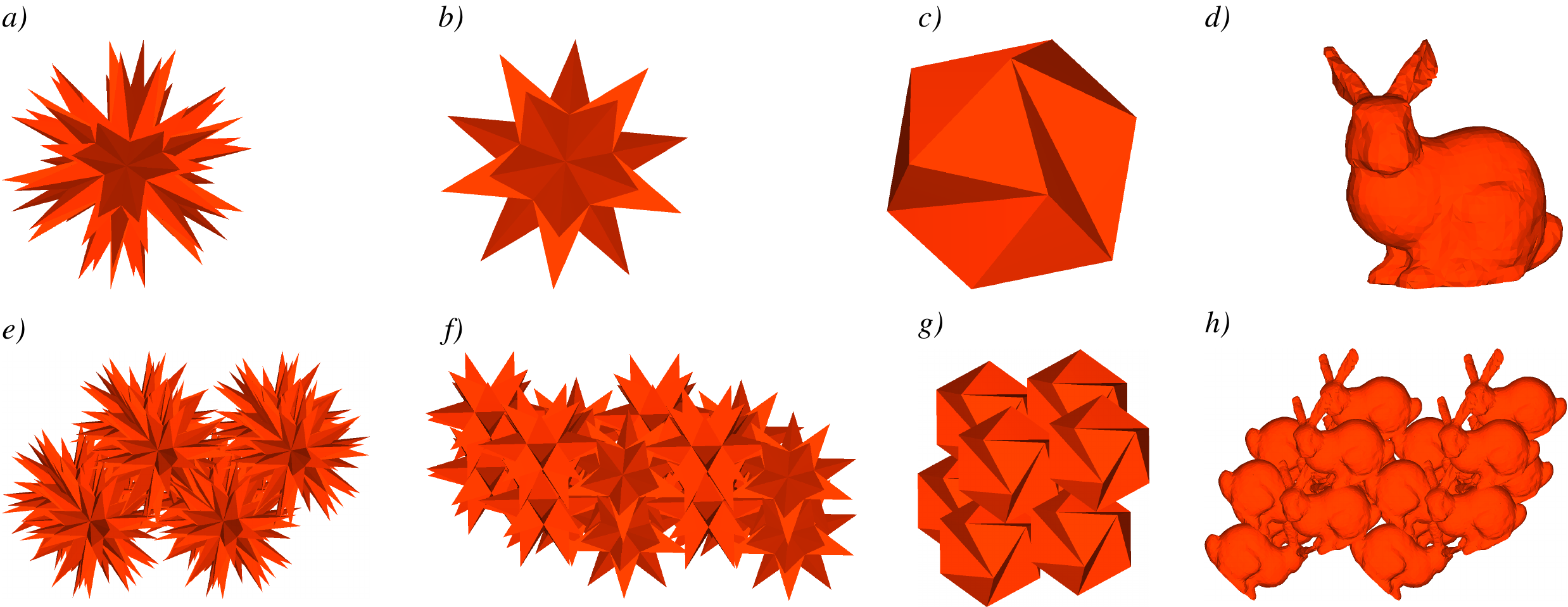}
\caption{\label{fig:nonconv} \textbf{Four non-convex particle species and their densest-packed crystals.} An Echidnahedron (a), a Great Stellated Dodecahedron (b), Jessen's Orthogonal Icosahedron (c), and a Stanford Bunny model (d). The crystal structures found by FBMC are given in (e) - (h) respectively, which show the unit cell and 7 of its periodic images. From left to right the packing fractions are $\phi_{\mathrm{lb}} = 0.294$, $0.889$, $0.749$, and $ 0.661$, for $N = 1$, $2$, $1$, and $2$ particles in the unit cell, respectively. Note the high degree of interpenetration for the Great Stellated Dodecahedron structure, which achieves a surprisingly high packing fraction.}
\end{figure*}

We have also applied our method to study non-convex (irregular) shapes, which may even contain holes, thereby going beyond existing studies. Fig.~\ref{fig:nonconv} shows representations of the shape and predicted crystal structure for 4 different particles. Additional information on these and nine other non-convex bodies can be found in Ref.~\cite{EPAPS}. These candidate crystal structures can be used in theoretical approaches or in  simulations with larger system sizes to determine their stability using, e.g., free-energy calculations. Non-convex particles can pack remarkably densely, even when they are not a priori designed to do so, as in the case of Escher's Solid. A total of 12 out of the 13 shapes achieve a regular packing with $\phi_{\mathrm{lb}} > 0.45$, and 10 of them achieve $\phi_{\mathrm{lb}} > 0.55$. The ISA upper bound is not very strong in these cases, since we obtain $\phi_{\mathrm{ub}}^{\circ} = 1$ for all but one of the models, due to their high asphericity. It is also clear from our result that the packing of non-convex (irregular) objects does not possess an intrinsic hierarchy, e.g., in terms of asphericity and symmetry.

A large fraction of the shapes we have considered is relatively simple, i.e., the tessellation of the particles only involves several dozen triangles. The simulations for such particles typically take between $10$ minutes and $2$ hours on a modern 2.0 GHz desktop PC. However, by the general applicability of triangular tessellation, it is also possible to study far more complex models. Two such models are the Hammerhead Shark, see Ref.~\cite{EPAPS}, and the Stanford Bunny, see Fig.~\ref{fig:nonconv}d and~\ref{fig:nonconv}h, which contain 5,116 and 3,756 triangles, respectively. Even for these high-triangle models the method proves to be relatively quick, the total run length of the simulations did not exceed $175$ hours.

\begin{figure*}[!htb]
\includegraphics[scale=0.60]{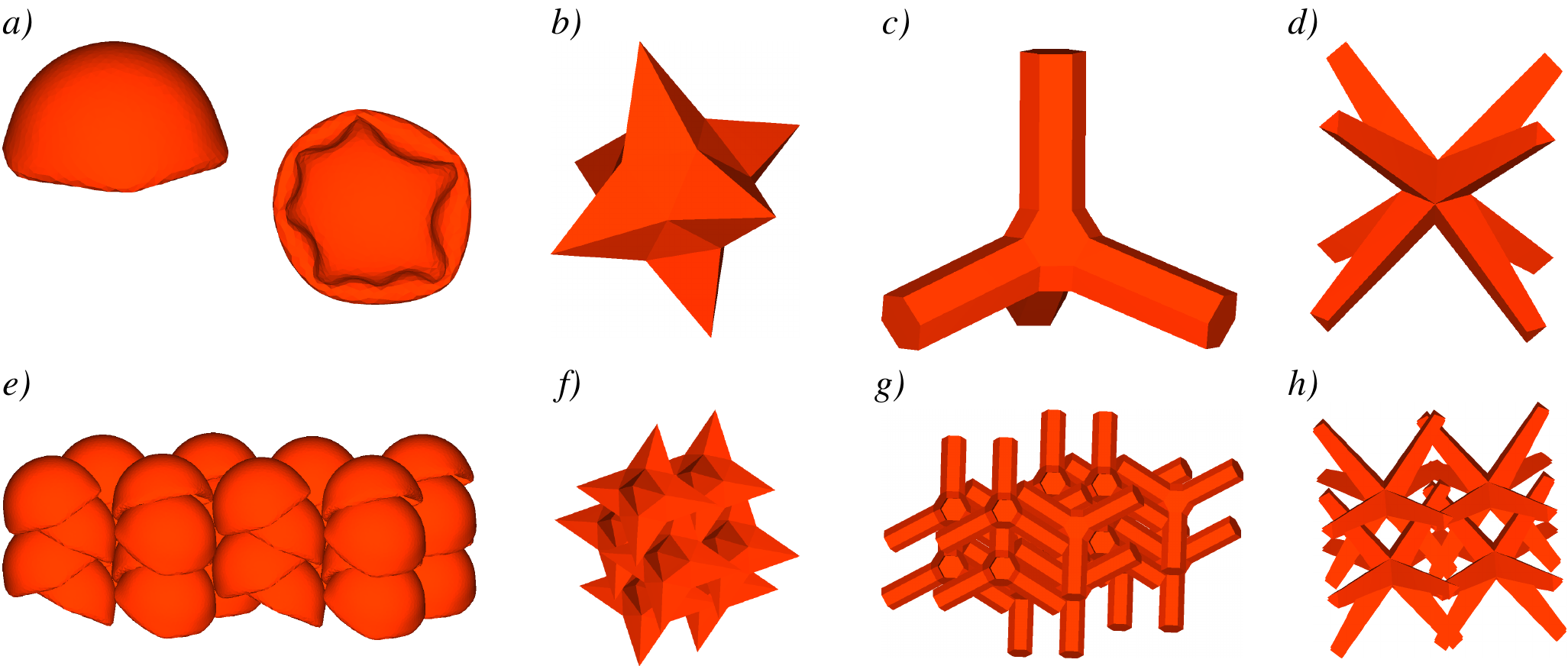}
\caption{\label{fig:realistic} \textbf{Four models of particles which can be synthesized and their densest packing.} (a) Two views of a Cap, (b) a Nanostar, (c) a Tetrapod, and (d) an Octapod. The structures of the densest-known regular packing of such particles are given in (e) - (h) respectively, showing the unit cell and 7 of its periodic images.}
\end{figure*}

The FBMC technique is also perfectly suited to study nanoparticle and colloidal systems. As examples we consider several non-convex shapes of experimentally available particles: Caps~\cite{Quilliet}, Nanostars~\cite{Zhao}, Tetrapods~\cite{Manna1}, and Octapods~\cite{Manna0}. Fig.~\ref{fig:realistic} shows representations of these particles and some of their crystal structures. Here, we focus on the excluded volume interactions due to the complex particle shape and neglect any other interactions such as, Coulomb, magnetic, and VanderWaals forces. However, any additional soft interaction term required to accurately describe an experimental system can easily be included. Even particle deformation is in principle possible, but its implementation into the FBMC scheme goes beyond the scope of our investigation.

One physical system we consider in more detail is that of the colloidal Cap. The shape of this Cap model, see Fig.~\ref{fig:realistic}a, is derived from a numerical analysis of the collapse of a spherical shell with bending and in-plane stretching elasticity submitted to an external isotropic pressure~\cite{Quilliet}. The model we use here is therefore more representative of an actual cap than the idealized model of a bowl in Ref.~\rcite{Marechal}. However, the present Cap model yields crystal structures similar to the ones described in Ref.~\rcite{Marechal}, namely: columnar, braided, and inverse braided~\cite{EPAPS}. Fig.~\ref{fig:realistic}e shows a braided configuration. Support is thus provided for the idea that the bowl shape captures the essential shape-related physics of these colloidal Cap systems. It should be noted that our simulations are much more computationally expensive than those of Ref.~\rcite{Marechal}, since the Cap tessellation contains 3,850 triangles in order to obtain a reasonably smooth approximation to the actual shape. Typical simulation times for Ref.~\rcite{Marechal} are in the order of $10$ minutes, whereas ours are around $150$ hours. Nevertheless, simulations of such complex shapes can be performed with the intent to verify the applicability of results obtained for simpler models. For faceted nanoparticle and colloid shapes~\cite{Solomon_Glotzer}, as well as smooth objects approximated by a lower number of triangles, the simulation times are not prohibitive; Nanostars, Octapods, and Tetrapods take the order of $30$ minutes to simulate.

In conclusion, we have developed a novel method to study  systems consisting of irregular, non-convex, and punctured objects. We have employed this technique to determine the densest regular packing and to predict candidate crystal structures in a rigorous way. The complex problem of the packing of such shapes has been reduced to determining a suitable approximation of a given particle in terms of triangles. Using this technique, we have predicted candidate crystal structures for non-convex particles and we have improved upon the literature values for the densest packings of convex solids. In addition, our method can easily be extended to study dense amorphous (granular) and quasicrystalline packings and systems of arbitrarily shaped colloids and nanoparticles in- and out-of-equilibrium. Simulations can be performed with or without external fields, e.g., gravitational or electric in nature, complex inter-particle interaction potentials, or even particle shape deformation. Our method thus opens the way to a more comprehensive study of the material and structure properties than has previously been considered feasible.

M.D. acknowledges financial support by a ``Nederlandse Organisatie voor Wetenschappelijk Onderzoek'' (NWO) Vici Grant, and R.v.R. by the Utrecht University High Potential Programme.

\end{document}